\documentclass[10pt]{article}
\usepackage[OE]{express}

\usepackage{floatrow}

\usepackage[greek,english]{babel} 
\usepackage[T1]{fontenc}
\usepackage[utf8]{inputenc}
\usepackage{upgreek} 
\usepackage{lmodern} 
\usepackage{subcaption}

\usepackage{epsfig}
\usepackage{amssymb}
\usepackage{mathrsfs}
\usepackage{times}
\usepackage{color}
\usepackage{ifpdf} 
\usepackage{graphics}
\ifpdf
\usepackage{epstopdf}   
\usepackage{url}
\fi
\usepackage{graphicx} 
\usepackage{amsmath}
\usepackage{amsfonts}
\usepackage{siunitx}

\newcommand{\bra}[1]{\ensuremath{\left\langle#1\right|}}
\newcommand{\ket}[1]{\ensuremath{\left|#1\right\rangle}}

\newcommand{\bibo}{BiB$_3$O$_6$ }

\usepackage[colorinlistoftodos,prependcaption,textsize=tiny,textwidth=3.5cm]{todonotes}

\begin{document}
	
	\title{High production rate of single-photon and two-photon Fock states for quantum state engineering}
	
	\author{Martin Bouillard,\authormark{1} Guillaume Boucher,\authormark{1} J\'ulia Ferrer Ortas,\authormark{1} Bhaskar Kanseri,\authormark{1,2} and Rosa Tualle-Brouri\authormark{1,*}}
	\address{\authormark{1}Laboratoire Charles Fabry, Institut d'Optique Graduate School, Universit\'e Paris-Saclay, 2 Avenue Augustin Fresnel, 91127, Palaiseau, FRANCE\\
	\authormark{2}Experimental Quantum Interferometry and Polarization (EQUIP), Department of Physics, Indian Institute of Technology Delhi, Hauz Khas, New Delhi-110016 India
}
	
	\email{\authormark{*}rosa.tualle-brouri@institutoptique.fr} 
	
	

	
	\begin{abstract}

		We report the implementation of a high-rate source of single and two-photon states. By combining the advantages of short pulses and cavities, heralding rates up to 250~kHz with 88\% fidelity have been obtained for the single photons as well as 800~Hz with 82\% fidelity for the two-photon states. Furthermore, we developed a setup where the homodyne measurement is conditioned by the heralding of the quantum states, enabling the detection of most of the heralded events. This allows a faster characterization of the photon source leading to an increase in the fidelities up to 91\% and 88\% respectively for the single-photon and two-photon Fock states. Such high rates and fidelities in the generation of elementary Fock states may open the path for the production of complex quantum states. 
	\end{abstract}
	

	\bibliographystyle{osajnl} 
	\bibliography{Biblio_FS} 
	
	\section{Introduction}
	
	Fock states are one of the most elementary resources towards quantum information and communication protocols \cite{KnillLaflammeMilburn2001,GisinThew2007,Kimble2008a,ObrienFurusawaVuckovic2009,PanChenLuEtAl2012,EtesseBlandinoKanseriEtAl2014,EtesseKanseriTualle-Brouri2014,AndersenNeergaard-NielsenLoockEtAl2015}. Over the last decades, important achievements have been seen for their generation \cite{OurjoumtsevTualle-BrouriGrangier2006,CooperWrightSoellerEtAl2013,SapienzaDavancoBadolatoEtAl2015,MaHartmannBaldwinEtAl2015,NgahAlibartLabonteEtAl2015,JeannicVermaCavaillesEtAl2016,SomaschiGieszSantisEtAl2016,LvovskyHansenAicheleEtAl2001}. Devices such as quantum dots in cavities are approaching a deterministic generation of single photons but their non-unity efficiency, in addition to their non-heralding nature, lead to a low fidelity which makes them hard to use, especially for continuous-variables protocols. On the opposite side, sources based on spontaneous parametric down conversion (SPDC) produce high-fidelity single-photon states in a probabilistic fashion. The high fidelities of the created states and the ease of implementation have made them highly popular. Another advantage of SPDC is that it can also produce multiple-photon Fock states, though the heralding rate drops quickly with the number of photons \cite{CooperWrightSoellerEtAl2013,WangChenLiEtAl2016}. The probabilistic nature of SPDC can be managed with the implementation of quantum memories: different setups have been developed over the years using atoms, bulk crystals or optical cavities \cite{YoshikawaMakinoKurataEtAl2013,JobezTimoneyLaplaneEtAl2016,KanedaXuChapmanEtAl2017}, making the heralding protocols a very promising tool towards quantum communication and computation.

	A way to characterize the quality of the Fock states emitted by a source is to analyze the number of photons available within some time interval \cite{PomaricoSanguinettiGuerreiroEtAl2012,JinShimizuMorohashiEtAl2014,SapienzaDavancoBadolatoEtAl2015,MaHartmannBaldwinEtAl2015,NgahAlibartLabonteEtAl2015,SomaschiGieszSantisEtAl2016}. The visibility in quantum interferences, like in Hong-Ou-Mandel experiments \cite{HongOuMandel1987,JinShimizuMorohashiEtAl2014,SomaschiGieszSantisEtAl2016}, can also be introduced, characterizing the ability to emit photons in optical modes that overlap well on a beam-splitter. However many protocols require Fock states in a well-defined spatio-temporal optical mode: this is especially the case when continuous variables are implied, where modes have to match the local oscillator in order to perform quadrature measurements with a homodyne detection~\cite{LvovskyHansenAicheleEtAl2001,OurjoumtsevTualle-BrouriGrangier2006,CooperWrightSoellerEtAl2013,YoshikawaMakinoKurataEtAl2013,EtesseBouillardKanseriEtAl2015}. Sources devoted to such applications therefore have to satisfy more stringent conditions. In this paper, we report the implementation of a source of Fock states using SPDC. By combining the advantages of pulses and cavities~\cite[and references therein]{Zavatta_2008_PRA_quantum, Krischek_2010_NP_Ultraviolet,KanseriBouillardTualle-Brouri2016}, we achieve a heralding rate of 250~kHz and 800~Hz, respectively for the generation of single and two-photon states. The quality of the produced states is measured using homodyne detection (HD). Using a maximum likelihood technique we reconstruct the density matrix of the input state and obtain a fidelity\footnote{$\mathcal{F}=\mathrm{tr}\left(\rho\ket{\psi}\bra{\psi}\right)$ with $\ket{\psi}$ the reference state and $\rho$ the experimental density matrix upon which the fidelity $\mathcal{F}$ is calculated~\cite{jozsa1994fidelity}.} as high as 88\% and 81\% for the single and two-photon states, respectively. To our knowledge, these are the highest emission rates and fidelities obtained so far in the pulsed regime for Fock states with HD characterization, representing a new step towards efficient quantum protocols with light.
	
	
	\section{Experimental setup}

	\begin{figure}[b]
		\centering
		\includegraphics[ width=\textwidth]{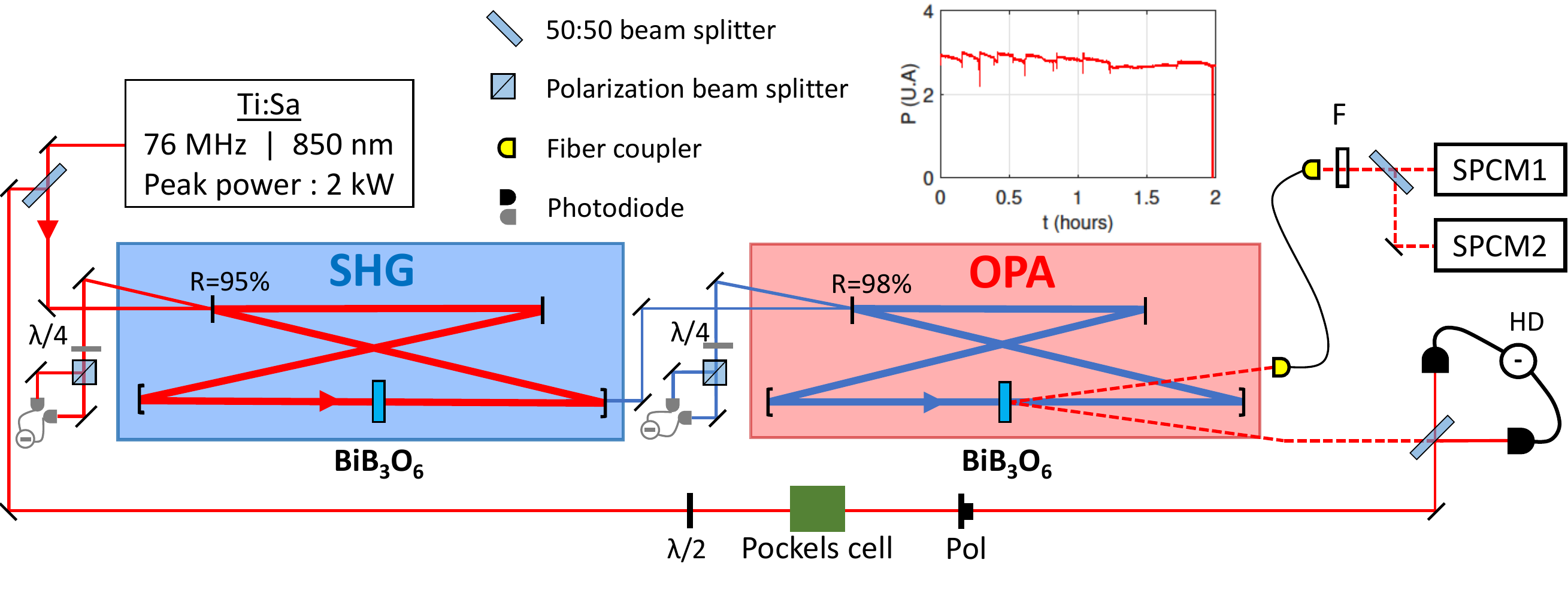}
		\caption{Experimental setup for the generation of single and two-photon states. SHG: second harmonic generation, OPA: optical parametric amplification, Pol: polarizer, HD: homodyne detection, F: spectral filter, SPCM: single photon counting module. The radius of curvature of the concave mirrors of both cavities is 1~m. Inset: stability of the OPA cavity. Jumps correspond either to an automatic relocking of the cavity, or to a manual adjustment of the cavity length. Apart from the input couplers, the reflectivity of the mirrors are R>99.9\% at 850~nm for the SHG cavity and 425~nm for the OPA cavity.}
		\label{fig:schemeexp}
	\end{figure}
	
	The setup is illustrated in figure \ref{fig:schemeexp}. In order to produce the quantum states, we use a pulsed Ti:~Sapphire laser with a repetition rate of 76~MHz. The output pulses have a central wavelength of 850~nm, a temporal length of 2.6~ps and a peak power of 2~kW. A part of the beam is extracted to serve as a local oscillator (LO). The main beam is sent to a second harmonic generation (SHG) stage in order to double the frequency of the laser. This SHG block is followed by an optical parametric amplification (OPA) stage where the Fock states are produced.
	
	
	To perform the SHG, we send the 850~nm pulses into a type I \bibo crystal (2-mm-thick) to produce pulses at 425~nm. In order to enhance the conversion efficiency, the crystal is placed in a bow-tie cavity of finesse 50~\cite{KanseriBouillardTualle-Brouri2016,cruz2007external,Takida2011}. As the length of the cavity has to match the distance between two consecutive pulses, the cavity is 3.95~m long. The crystal is installed between the two concave mirrors where the waist of the beam is $\sim$~200~$\upmu$m. In order to reduce optical losses, both sides of the crystal have been treated with anti-reflection coatings at 425 nm and 850 nm. The phase of the cavity is locked using the Hänsch-Couillaud technique \cite{HanschCouillaud1980,KanseriBouillardTualle-Brouri2016}. We reach a conversion efficiency up to 70\% using this approach, leading to a peak power of 1.1~kW (for an input power of 1.6~kW).
	
	
	The frequency doubled beam is then directed to another 3.95-m-long bow-tie cavity, namely the OPA cavity, in order to produce pairs of photons via SPDC. A 1-mm-thick type-I \bibo crystal is placed between the two concave mirrors of the cavity, where the beam waist is approximately 100~$\upmu$m. The locking scheme is the same as for the SHG cavity. The cavity enhances the peak power of the frequency-doubled pump beam by a factor $\sim$~50, leading to a peak power of~$\sim$~50~kW. A half-wave plate followed by a polarizer placed ahead of the cavity allows to control the polarization of the beam as well as the power resonating inside the cavity. The \bibo crystal is used in a non-collinear configuration by slightly tilting the crystal (12') from the collinear configuration. The signal beam is first coupled into a 1-m-long polarization-maintaining single-mode fiber, allowing a spatial filtering of the single photon. Then, as the SPDC process is frequency-degenerate, a spectral filtering is achieved by combining a grating with a slit placed in the focal plane of a lens. This technique allows the tuning of the central frequency of the single photons in order to match the central wavelength of the laser. The bandwidth of the filtering is optimized in order to remove as much frequency correlations as possible while maintaining a strong heralding rate. Following the spectral filtering stage, we probabilistically separate the photons using a 50:50 beam-splitter. Two single-photon counting modules (SPCM, Perkin Elmer SPCM-AQR-13) placed in the output ports of the splitter, allow to measure  single-photon Fock states, when only one SPCM clicks, and two-photon Fock states, when both SPCMs click for the same pulse. The overall efficiency of the detection is measured to be around 6\%.

	\captionsetup[subfigure]{position=top, labelfont=normalfont,textfont=normalfont,singlelinecheck=off,justification=raggedright}
	\begin{figure}[t]
		\centering
		\subcaptionbox{\label{sfig:Fid_vs_Pf}}{\includegraphics[height=0.45\textwidth]{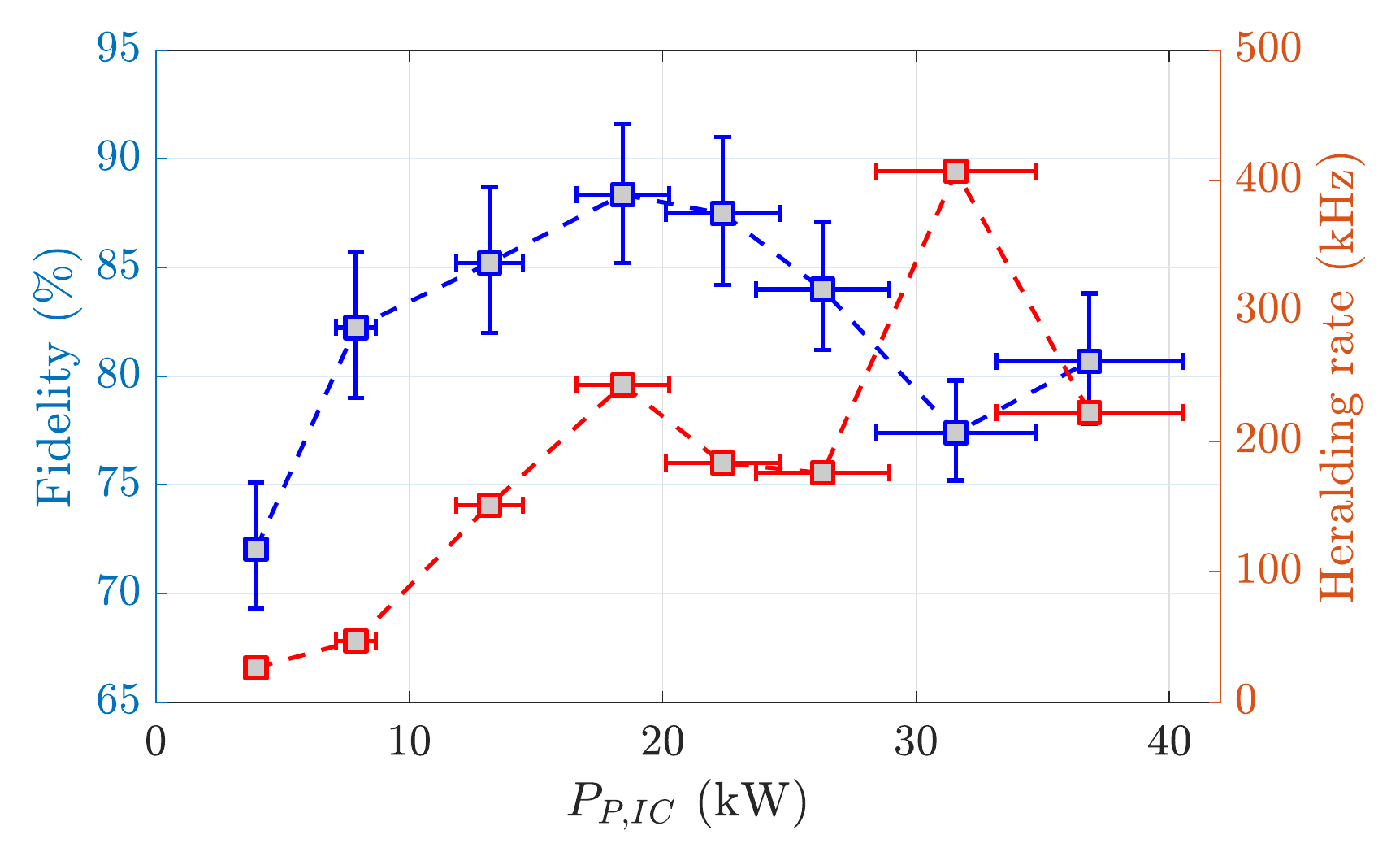}}\\
		\subcaptionbox{\label{sfig:Distribution_SP}}{\includegraphics[height=0.42\textwidth]{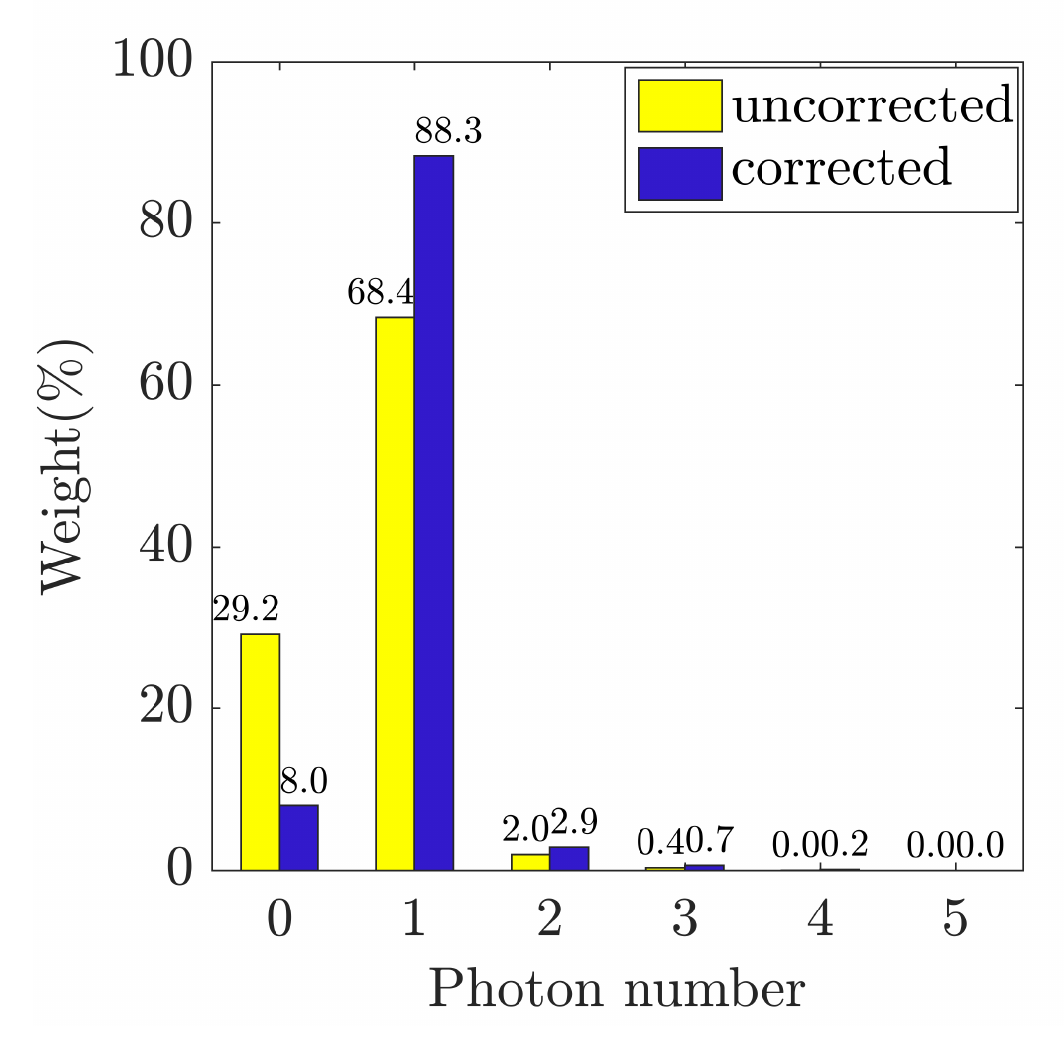}}
		\subcaptionbox{\label{sfig:WF_SP}}{\includegraphics[height=0.42\textwidth]{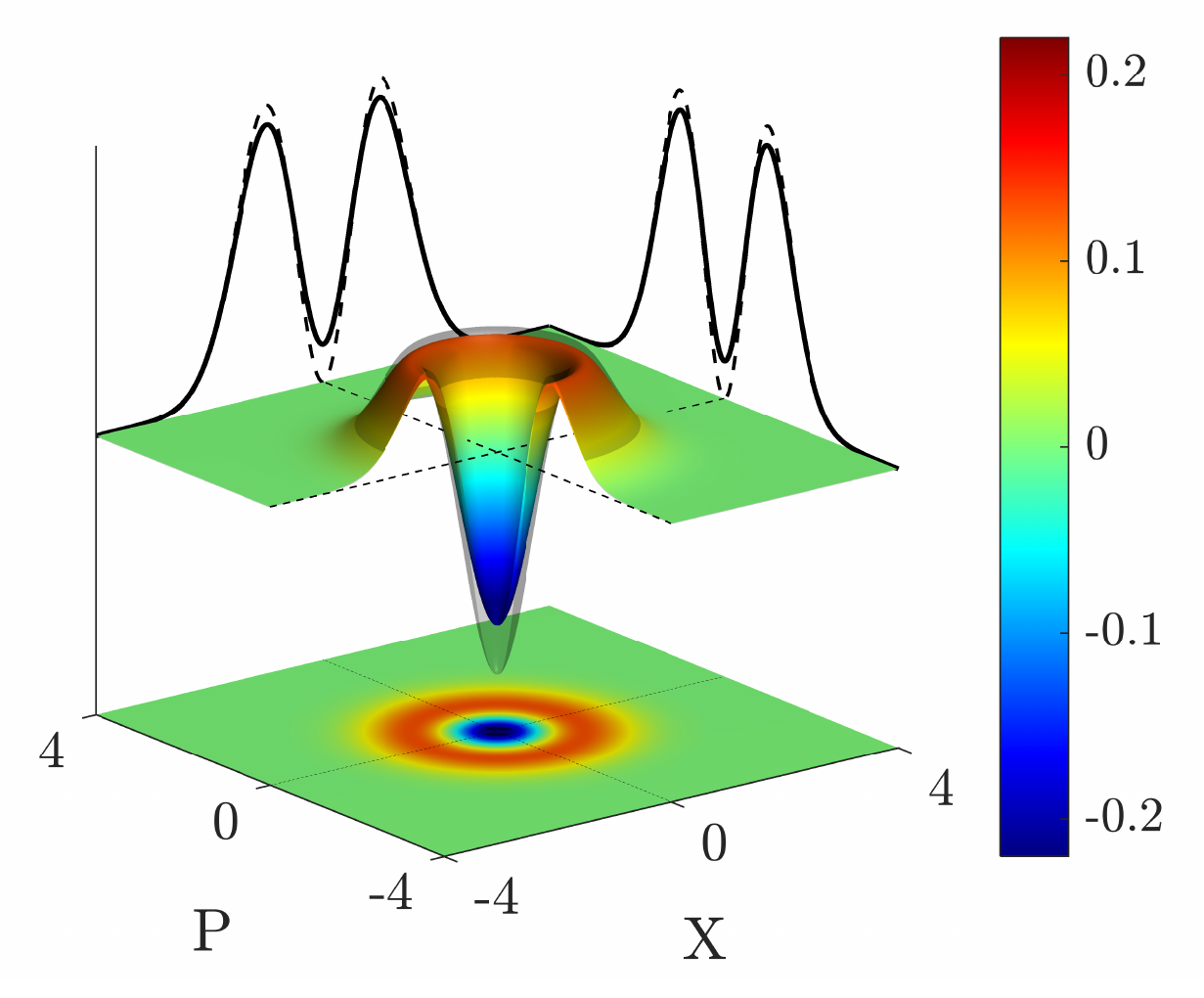}}	
		\caption{Experimental results. (a) Fidelity and heralding rate as functions of the IC pump peak power. The fidelity reaches 88\% for a heralding rate of 250~kHz at a pump peak power of  $\sim$~20~kW. (b) Diagonal elements of the density matrix without and with correction of the detection losses. (c) Colored: Wigner function of the measured state; dark shade: Wigner function of the theoretical single-photon Fock state; solid lines: projection of the obtained state along one axis; dashed lines: distributions of a theoretical single-photon Fock state.}  
		\label{SP}
	\end{figure}

	The heralded states are then characterized by quantum state tomography using a HD~\cite{Hansen01}. The overall detection efficiency of the HD is $\upeta_{\mathrm{HD}}$=$\upeta_{\mathrm{PD}}\upeta_\mathrm{C}=$76$\pm$3\% (where $\upeta_{\mathrm{PD}}$=94$\pm$2\% is the quantum efficiency of the photodiodes, and $\upeta_{\mathrm{C}}$=C$^2$=81$\pm$1\% is the mode-matching efficiency, with C the contrast measured between the idler beam and the local oscillator.
 As the estimation of detection losses leads to the main source of error, we estimate the error bars of the fidelity by reconstructing the density matrix for 73\% and 79\% detection efficiency. It is worth noting that if the error on the contrast is statistical, the error on the quantum efficiencies of the photodiodes, due to the calibrations of the powermeters used to characterize them, is systematic. The HD has a few-MHz bandwidth, which is much lower than the repetition rate of the laser. Thus, we reduce the repetition rate of the laser using a Pockels cell (PC) on the local oscillator. The PC allows to switch the polarization of the pulses and, combined with a half-wave plate (HWP) and a polarizer, reduces the repetition rate of the local oscillator from 76 MHz to 1 MHz, allowing a complete characterization of the states emitted by our source.
	
	\section{Experimental results}

	In order to characterize our source, we acquire single-photon measurements for different intra-cavity (IC) peak powers. The IC peak power is determined by measuring a leak from one of the curved mirrors of the cavity. For each power, we acquire 50~000 single-photon data points with the HD. The single-photon data points are acquired when only one APD is triggered. We retrieve the Wigner function of the state by using a maximum likelihood technique \cite{Lvovsky2004}, where the maximum number of photons for the reconstruction algorithm is truncated to 5. As the HD is not part of the creation protocol, the losses of the HD can be corrected \cite{KissHerzogLeonhardt1995} allowing to reconstruct the quantum state actually heralded by the SPCMs before the HD. 
	
	\begin{figure}
		\centering
		
		\subcaptionbox{\label{sfig:Distribution_2P}}{\includegraphics[height=0.42\textwidth]{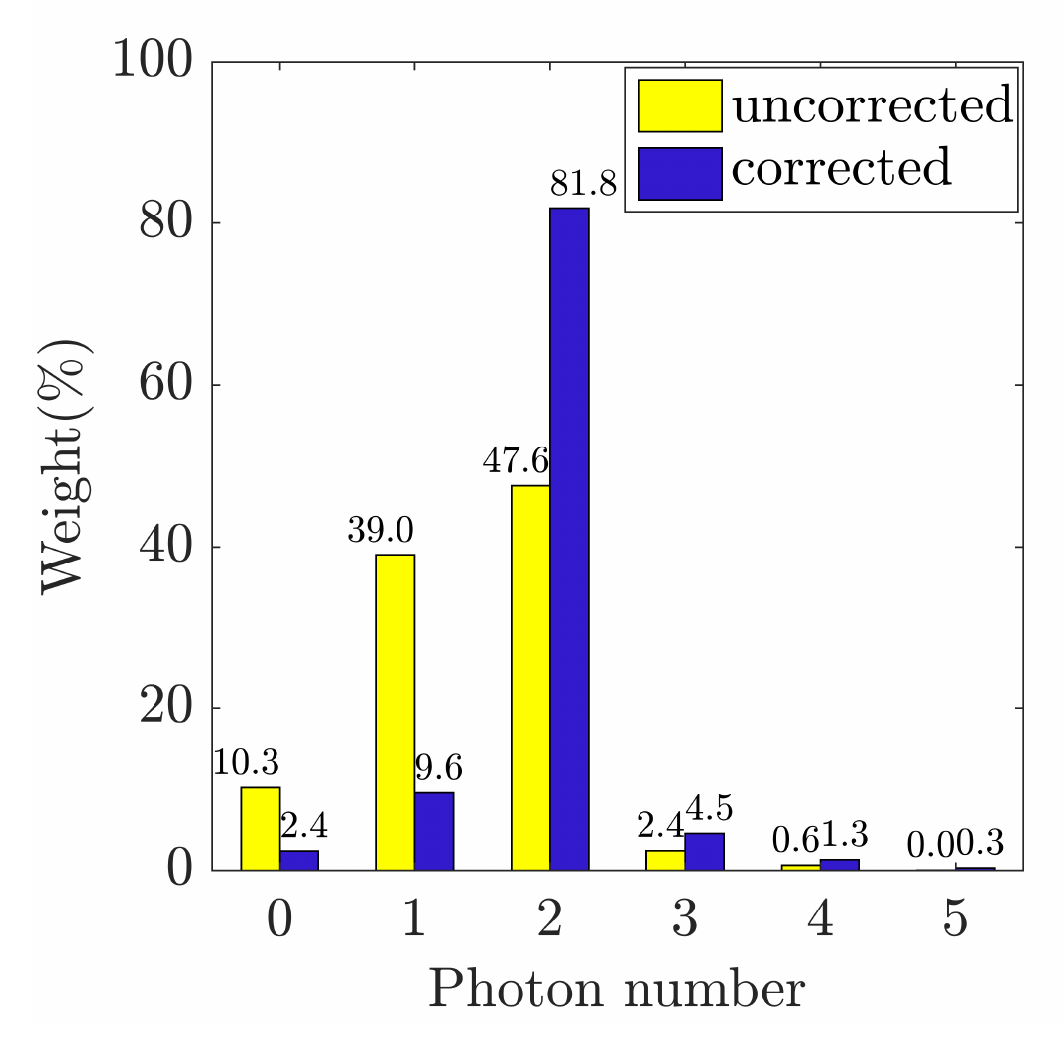}}
		\subcaptionbox{\label{sfig:WF_2P}}{\includegraphics[height=0.42\textwidth]{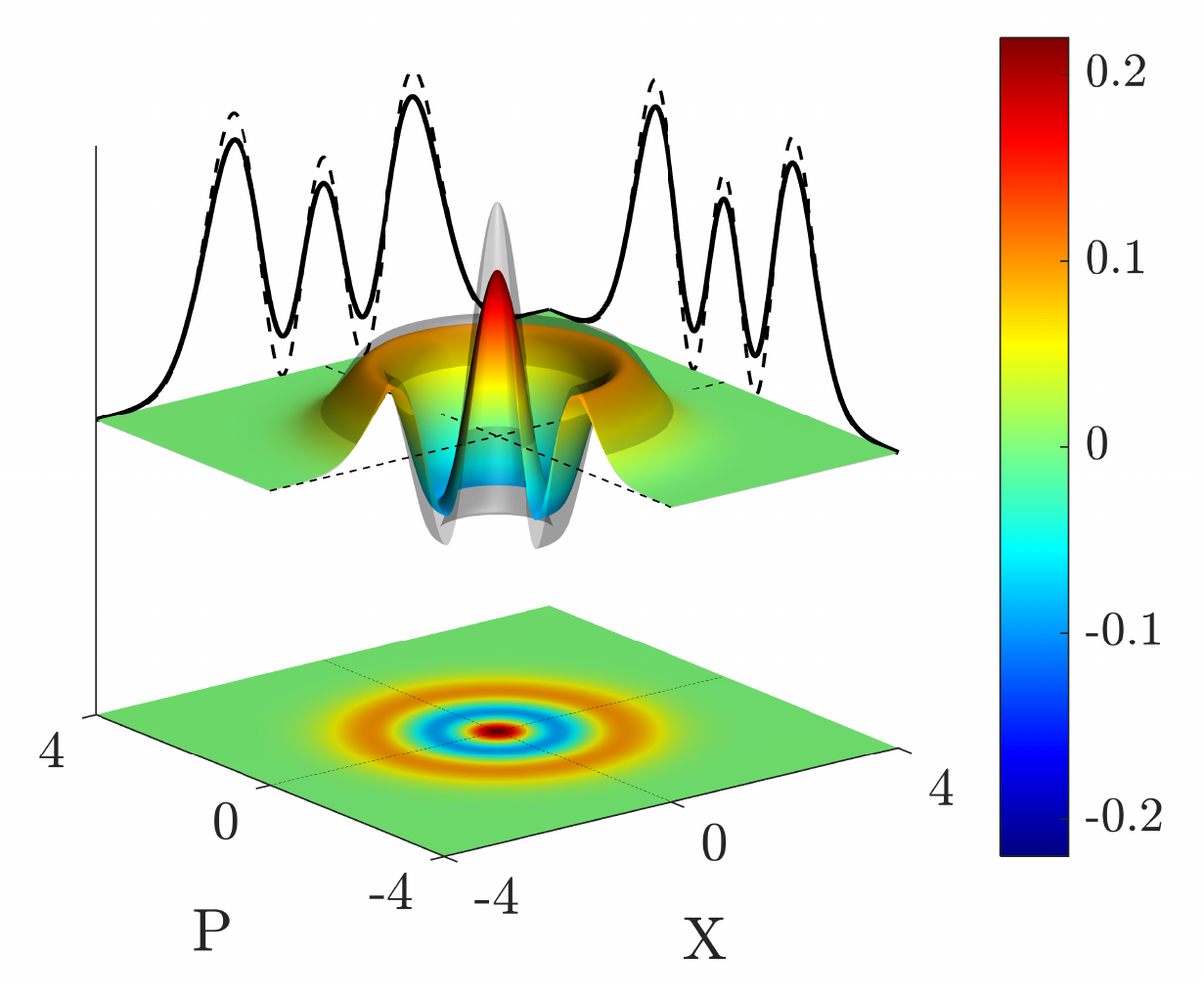}}
		\caption{Experimental results. (a) Diagonal elements of the density matrix without and with correction of the detection losses. (b) Colored: Wigner function of the measured state; dark shade: Wigner function of the theoretical single-photon Fock state; plain lines: projection of the obtained state along one axis; dashed lines: distributions of a theoretical two-photon Fock state.}  
		\label{2P}
	\end{figure}
	
	The figure \ref{SP}(a) represents the fidelity of the single-photons as a function of the IC peak power in the OPA cavity. We reach a fidelity of 88\% for an IC peak power of $\sim$~20~kW. On this figure, we note that the fidelity increases with the power, explained by the ratio of dark counts over single photons counts decreasing with the increase of power and thus heralding rate. Above an IC peak power of  $\sim$~20~kW, the fidelity starts to decrease. This behavior has two different origins : 
	First, thermal effects inside the crystal are appearing as we increase the IC power, reducing the quality of the photons. Second, the number of two-photon state events increases with the IC power.
	The diagonal elements for the obtained single-photon state, with and without error correction, for an IC peak power of 20~kW are presented in figure \ref{SP}(b), whereas the corresponding Wigner function (with error correction) is demonstrated in the figure \ref{SP}(c). 
	The solid lines and dashed lines represent respectively the probability distribution of the measured state (corrected from the detection losses) and the theoretical state. A fidelity of 88$^{+4}_{-3}$\% has been reached for an heralding rate of $\sim$~250~kHz, while the analysis rate is $\sim$~500~Hz. Compared to our previous experiment \cite{EtesseBouillardKanseriEtAl2015}, we increase the fidelity by 11\%, while we enhance the heralding rate by more than two orders of magnitude.


	We now look at the quality of the two-photon Fock states (i.e. when both SPCMs heralded a photon for the same pulse). Due to a low rate of analysis of few Hz, we acquire only 10~000~points. The diagonal elements and the Wigner function of the state, obtained for an IC peak power of 23~kW, are reconstructed via the maximum-likelihood technique with error correction (figures \ref{2P}(a) and \ref{2P}(b)). We achieve a fidelity of 82$\pm$6\% for a heralding rate of 800~Hz. Both values, the heralding rate and the fidelity are, to our knowledge, the highest for the production of two-photon states using continuous variables.

	
	\section{An optical delay line for improved analysis rates}

	In the previous sections, a Pockels cell is used to slow the repetition rate of the LO down to 1~MHz, in order to avoid saturation of the HD, meaning that all the events are not analyzed. Since the photons are produced randomly and since we collect a sufficient amount of statistics, we can be confident that our reconstruction yields a faithful representation of the heralded state. The ability to measure all the events concerns the detection system, and is disconnected from the production part. However, the bandwidth of the HD is a limitation that one has to manage, and the present section is dedicated to overcome this issue. 

A fast HD~\cite{CooperWrightSoellerEtAl2013} could solve this problem, but the benefit in terms of bandwidth can be mitigated by additional electronic noise and a loss of detection efficiency. We propose another approach: while the HD cannot support the 76~MHz emission rate of the laser, it would have no difficulty to follow the sub-MHz heralding rate of the Fock-state source. One just has to trigger the Pockels cell with the photons heralding events. This will raise the problem of the delay required by the electronics for the triggering process, but this problem can be solved with an optical delay line inserted between the Fock states source and the HD. 

We developed an optical delay line based on bouncing the idler beam between a plane and two spherical (2.5~m curvature radius) high reflectivity mirrors (>99.99\%), separated by 4.8~m. With the control of the round trips number, it allows to tune the delay of the heralded Fock states from~32 to~288~ns by steps of 32~ns with optical losses lower than 1\%.

\begin{figure}
	\centering
	
	\subcaptionbox{\label{sfig:Distribution_SP_QMC}}{\includegraphics[height=0.42\textwidth]{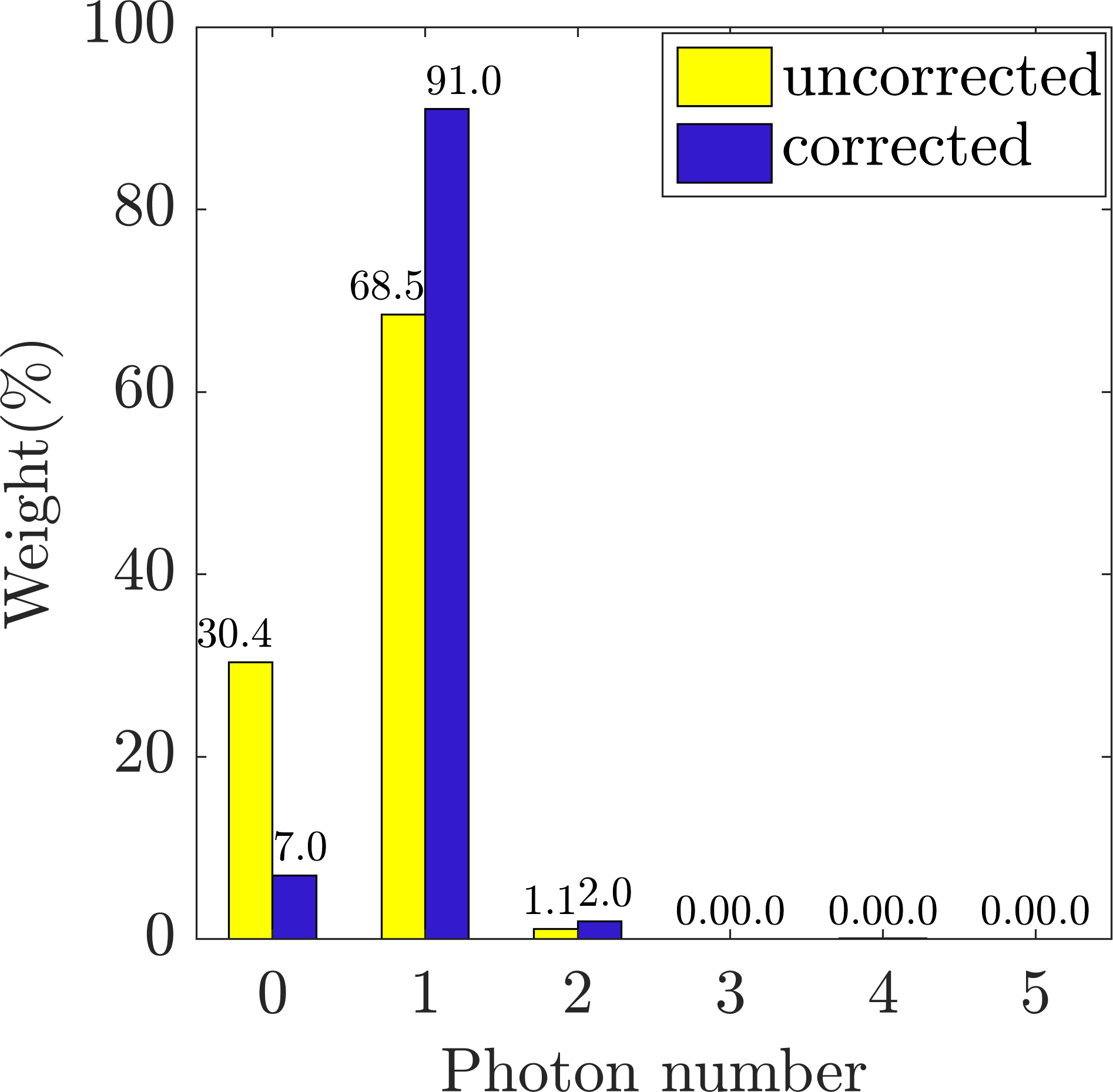}}
	\subcaptionbox{\label{sfig:WF_SP_QMC}}{\includegraphics[height=0.42\textwidth]{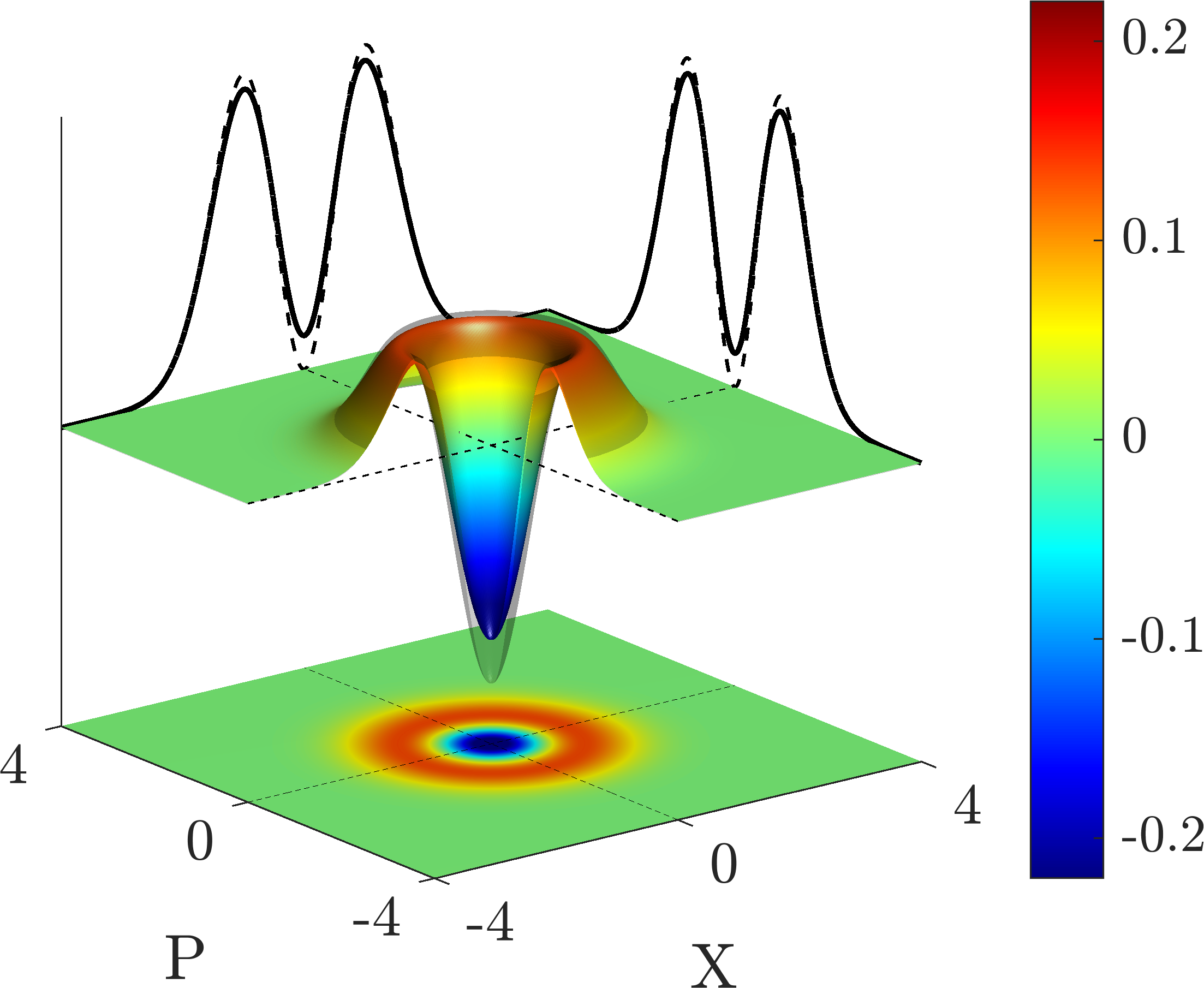}}
		
	\subcaptionbox{\label{sfig:Distribution_2P_QMC}}{\includegraphics[height=0.42\textwidth]{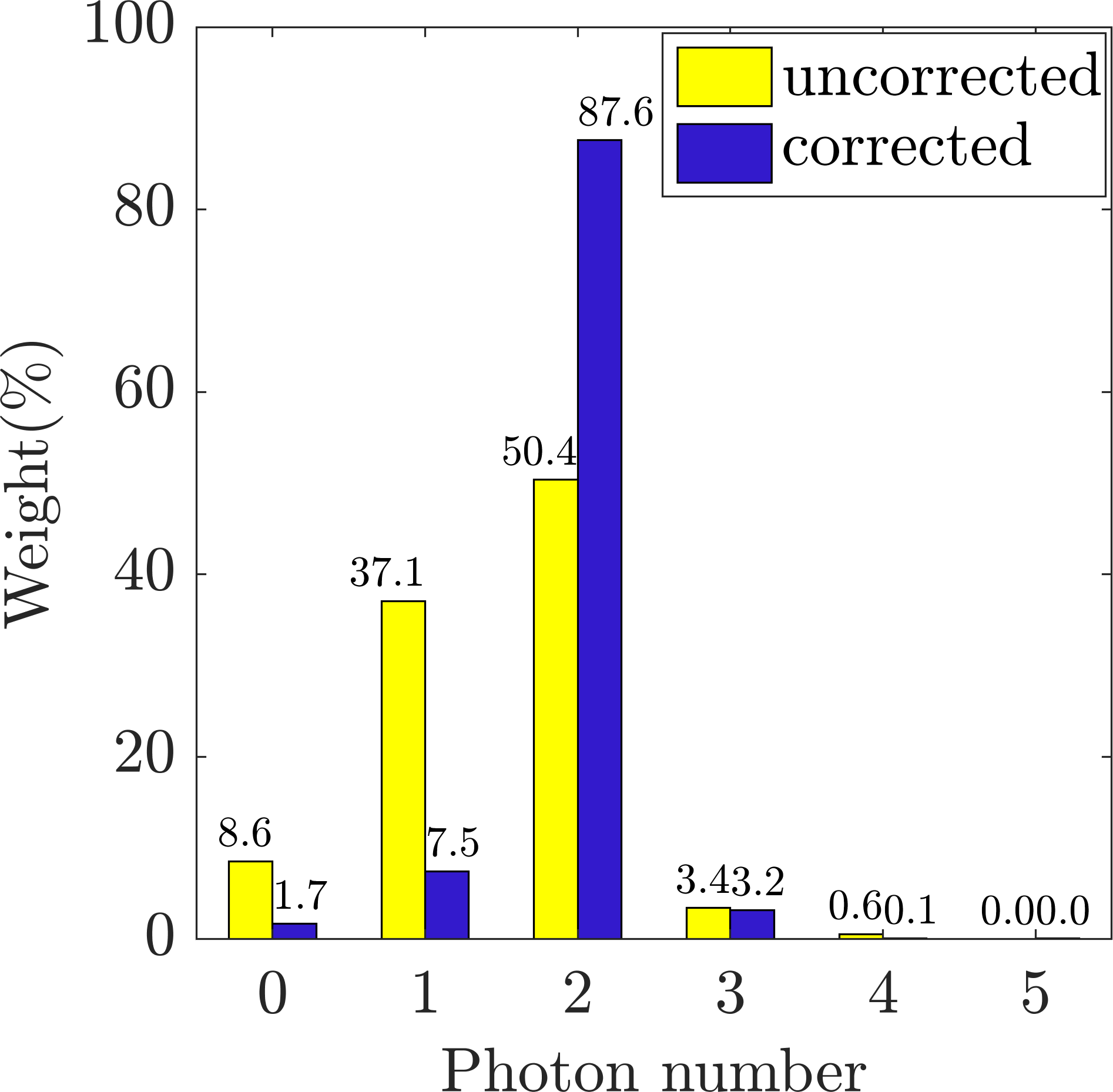}}
	\subcaptionbox{\label{sfig:WF_2P_QMC}}{\includegraphics[height=0.42\textwidth]{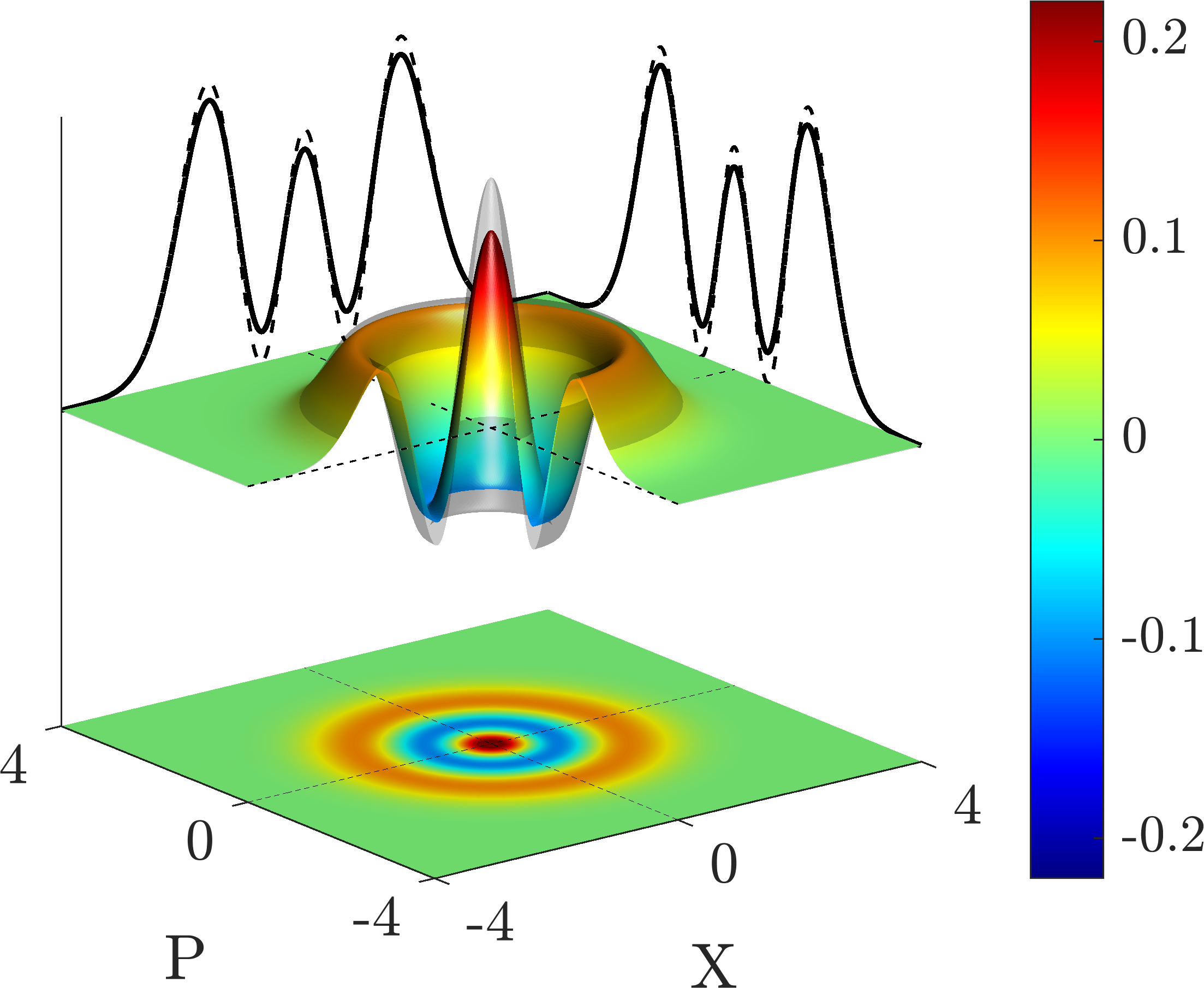}}
	\caption{Diagonal elements of the density matrix for the single-photon (a)~and two-photon (c)~Fock states obtained with an optical delay line for improved detection rates, without and with correction of the detection losses. (b)~and (d)~Wigner functions of the reconstructed single-photon (b)~and two-photon (d)~Fock states; dark shade: Wigner function of the theoretical states; plain lines: projections of the obtained state along one axis; dashed lines: distributions of the theoretical states.}
	\label{fig:SP_QMC}
\end{figure}

With this system, it is possible to measure all the events, except a small part which is falling into the dead-time of the HD ($\sim\SI{1}{\micro\second}$). The experimental settings here differ a little bit from what was presented in the previous section but, with an analysis rate of $\sim$~160~kHz for the single-photon states, and of $\sim$~250~Hz for the two-photons states, we show that we can reach measurement rates that are very close to the corresponding heralding rates(respectively $\sim$~200~kHz and $\sim$~250~Hz in this experiment for the single and two-photons states). The rates of unmeasured events by the HD are here estimated to be of $\sim 40$~kHz ($< 1$~Hz) for the single photon (two-photon) Fock states. In addition, this detection modality does not induce any deterioration of the quality of the measured states. On the contrary, we observed an improvement with a corrected fidelity reaching 91±4\% and 88±4\% for the single and two-photon states, respectively. The tomographies of these states are presented on the figure~\ref{fig:SP_QMC}.


	\section{Conclusion}
	We experimentally demonstrate a fast production of single and two-photon Fock states. The fidelity of the states reaches 88\% for the single photons with a heralding rate of 250~kHz, and~82\% with a heralding rate of 800~Hz for the two-photon state. With an optical delay line that allows to trigger the measurements with the detection events, we reach a detection rate of 160~kHz for single-photon states with a 91\% fidelity, and of 250~Hz for two-photons states with a 88\% fidelity. As Fock states are the first step for the production of a multitude of quantum states, such as the production of optical Schrödinger cat states~\cite{EtesseBouillardKanseriEtAl2015, HuangJeannicRuaudelEtAl2015, SychevUlanovPushkinaEtAl2017}, the production of those elementary quantum states with high fidelities and heralding rates is essential. Combined with an optical quantum memory currently in development, such experiments could pave the way towards continuous-variables and hybrid quantum-information protocols.

	\section*{Funding}
	
	The authors acknowledge support from the Région Ile de France DIM NANOK for the HAQI project, and from the Agence Nationale de la Recherche (ANR) for the SPOCQ project (project No. ANR-14-CE32-019).
	
	\section*{Acknowledgments}
	The authors thank Imam Usmani and Rajiv Boddeda for fruitful discussions.
	
	\section*{Disclosures}
	The authors declare that there are no conflicts of interest related to this article.
	
\end{document}